\documentclass[prd,aps,twocolumn,amsfonts,showpacs,superscriptaddress,preprintnumbers,nofootinbib]{revtex4-2}
 
\usepackage{graphicx}
\usepackage[dvipsnames]{xcolor}
\usepackage{rotating}
\usepackage{bm,amsmath,amssymb}
\usepackage[utf8]{inputenc}
\usepackage{footmisc}
\usepackage{diagbox}
\usepackage{mathtools}    
\usepackage{dcolumn}
\usepackage{verbatim} 
\usepackage{bbold} 
\usepackage{slashed}
\usepackage[normalem]{ulem} 
\usepackage{tikz}
\usepackage{soul}

\usepackage[colorlinks=true, urlcolor=blue, linkcolor=blue, citecolor=blue, hyperindex=true, linktocpage=true]{hyperref}

\newcommand{\pars}[1]{\left(#1\right)}
\newcommand{\spars}[1]{\left[#1\right]}

\newcommand{\nn}{\nonumber\\}

\def\CN{\mathcal{N}}

\begin{document}

\title{Quantum origin of Ohm's reciprocity relation and its violation:\\ conductivity as inverse resistivity}

\author{Giorgio Frangi}
\affiliation{Higgs Centre for Theoretical Physics, University of Edinburgh, Edinburgh EH9 3FD, Scotland}

\author{Sa\v{s}o Grozdanov}
\affiliation{Higgs Centre for Theoretical Physics, University of Edinburgh, Edinburgh EH9 3FD, Scotland}
\affiliation{Faculty of Mathematics and Physics, University of Ljubljana, Jadranska ulica 19, SI-1000 Ljubljana, Slovenia}


\begin{abstract}
Conventional wisdom teaches us that the electrical conductivity in a material is the inverse of its resistivity. In this work, we show that when both of these transport coefficients are defined in linear response through the Kubo formulae as two-point correlators of conserved currents in quantum field theory, this Ohm's reciprocity relation is generically violated in theories with dynamical electromagnetism. We then elucidate how in certain special limits (e.g., in the DC limit in the presence of thermal effects, in certain 2+1$d$ conformal theories, and in holographic supersymmetric theories) the reciprocity relation is reinstated as an emergent property of conductive and resistive transport. We also show that if the response of a material is measured with respect to the total electric field that includes quantum corrections, then the reciprocity relation is satisfied by definition. However, in that case, the transport coefficients are given by the photon vacuum polarisation and not the correlators of conserved currents that dominate the hydrodynamic macroscopic late-time transport.
\end{abstract}

\maketitle

{\bf Introduction.---}Ohm's law is a cornerstone of physics. It  provides a simple framework to characterise materials that conduct electric currents. It is a phenomenological statement that stipulates that in a piece of a conductor, the voltage $V$ is linearly related to the total current $I$ flowing through the material: $V=R I$. The coefficient of proportionality $R$ is the material's resistance. A more fine-grained way to express Ohm's law is in the local form:
\begin{equation}\label{ohm_differential_Sigma}
    \mathbf{j}(\omega) = \sigma(\omega)  \mathbf E(\omega),
\end{equation}
where $\mathbf{j}(\omega)$ is the current density, $\sigma(\omega)$ the AC frequency-dependent electrical {\em conductivity} (matrix) and $\mathbf E(\omega)$ the electric field. To define the electrical resistivity, it is standard to invert Eq.~\eqref{ohm_differential_Sigma} and write
\begin{equation}\label{ohm_differential_Rho}
    \mathbf{E}(\omega) = \rho(\omega)  \mathbf j(\omega),
\end{equation}
where $\rho(\omega)$ is the electrical {\em resistivity} (matrix), simply defined as: $ \rho(\omega) = \sigma^{-1}(\omega)$ (see Refs.~\cite{Kittel,mahan}). In rotationally (and in 2+1$d$ also parity) invariant systems, $\sigma_{ij} = \sigma \delta_{ij}$ and $\rho_{ij} = \rho \delta_{ij}$. Hence,
\begin{equation}\label{Ohm_Reciprocity}
    \sigma(\omega) \rho(\omega) = 1.
\end{equation}
We will refer to Eq.~\eqref{Ohm_Reciprocity} as {\em Ohm's reciprocity relation}.

The simplest microscopic model of a conductor is the well-known Drude model (see e.g.~Ref.~\cite{mahan}). Despite its qualitative and quantitative value, it is a phenomenological setup that cannot be extended to the breadth of conductors of interest. Instead, more powerful techniques are required, which must ultimately account for the effects of the underlying quantum field theory (QFT). This can be done by using linear response theory, which connects a microscopic calculation of correlators from either kinetic theory, perturbative QFT or holography to the appropriate low-energy effective theory, such as hydrodynamics (see Refs.~\cite{Kovtun:2012rj,hartnoll2018holographic,zaanen2015holographic,KapustaGale,leBellac:thermalFT,chaikin_lubensky}). The macroscopic quantities such as $\sigma$ and $\rho$ are then identified with the infra-red (IR) limit of QFT correlators. In the process, a system is perturbed away from equilibrium by coupling an external field to a QFT operator. In a conductor, an external electric field ${\bf E}_{\rm ext}$ couples to the electric current, giving it a finite expectation value: $\langle {\bf j} \rangle \neq 0$. If the theory has dynamical electromagnetism, ${\bf E}_{\rm ext}$ couples to the dynamical electromagnetic field as well, giving $\langle {\bf E} \rangle \neq 0$. We write
\begin{equation}\label{Ohm_ext}
\langle \mathbf{j}(\omega) \rangle = \sigma(\omega)  \mathbf E_{\rm ext} (\omega), \quad \langle \mathbf{E}(\omega) \rangle = \rho(\omega)  \mathbf j_{\rm ext} (\omega),
\end{equation} 
where $\mathbf j_{\rm ext}$ is an external current related to $\mathbf E_{\rm ext}$ via Maxwell's equations, and the following Kubo formulae define the conductive transport coefficients:
\begin{equation}\label{Kubo_ext}
    \sigma(\omega) = \lim_{k\to0} \frac{G^{jj}_R (\omega,k) }{i\omega}, \quad \rho(\omega) = \lim_{k\to0} \frac{G^{EE}_R(\omega,k)}{i\omega}.
\end{equation}
Here, $k \equiv |{\bf k}|$ is the magnitude of the wavevector and $G_R$ denote the retarded two-point functions of ${\bf j}$ and ${\bf E}$, chosen to be longitudinal (parallel to ${\bf k}$). The Kubo formula for $\sigma$ is extremely well known. The formula for $\rho$, however, was to the best of our knowledge first derived in the context of magnetohydrodynamics formulated with higher-form symmetries in Ref.~\cite{Grozdanov:2016tdf} (see also \cite{Hernandez:2017mch,Armas:2018atq,Vardhan:2022wxz}).

An alternative definition of the conductivity that is sometimes used in the literature relates the expectation value of ${\bf j}$ and the total electromagnetic field ${\bf E_{\rm tot}} = {\bf E_{\rm ext}} + \langle {\bf E} \rangle$ consisting of the external source and the induced quantum field. In such case, 
\begin{equation}\label{Ohm_int}
\langle \mathbf{j}(\omega) \rangle = \tilde \sigma(\omega)  {\bf E_{\rm tot}}(\omega), \quad {\bf E_{\rm tot}}(\omega) = \tilde\rho(\omega)  \langle \mathbf j (\omega)\rangle.
\end{equation} 
The appropriate analogue Kubo formula for $\tilde \sigma$ is not given in terms of the conserved $U(1)$ current correlator but the retarded vacuum polarisation $\Pi_R$ (see e.g.~Ref.~\cite{mahan,leBellac:thermalFT}):
\begin{equation}\label{Kubo_int}
\tilde \sigma(\omega) =  \lim_{k\to 0} \frac{\Pi_R(\omega,k)}{i\omega} = \frac{1}{\tilde\rho(\omega)}.
\end{equation}
The validity of Eq.~\eqref{Ohm_Reciprocity} for $\tilde \sigma$ and $\tilde \rho$ follows from the definition \eqref{Ohm_int}. The vacuum polarisation $\Pi$ is the one-particle-irreducible (1PI) part of the photon Feynman (time-ordered) correlator $G_F^{AA}$, with $A^\mu$ parallel to ${\bf k}$. $\Pi_R$ is found from $\Pi$ after an analytic continuation.

It should be clear that, generically, $\sigma \neq \tilde \sigma$ and $\rho \neq \tilde \rho$ and that the use of either Eq.~\eqref{Ohm_ext} or \eqref{Ohm_int} must depend on the context of the physical scenario that the theory is attempting to describe. While the (intrinsically) defined conductive transport coefficients in terms of $\Pi_R$ can be computed in pertubative QFT, this is less clear non-perturbatively. Moreover, such coefficients are harder to discuss in an IR effective field theory context or an experimental setting, where it is the long-lived, conserved currents that play the central role in determining the dynamics of the low-energy degrees of freedom. Note that both expressions in Eq.~\eqref{Kubo_ext} can be understood as two-point correlators of conserved currents: the usual $j^\mu$ and the two-form current $J^{\mu\nu}$ related to ${\bf E}$ (see Ref.~\cite{Grozdanov:2016tdf}).

The aim of this paper is to show that, generically, Ohm's reciprocity relation \eqref{Ohm_Reciprocity}, as applied to the definitions using conserved currents in \eqref{Kubo_ext}, does {\em not} hold in theories with dynamical electromagnetism:   
\begin{equation}\label{Claim}
\sigma(\omega) \rho(\omega) \neq 1. 
\end{equation} 
While the claim \eqref{Claim} was initially made in Ref.~\cite{Grozdanov:2016tdf}, it is in this work that we explore and elucidate the difference between $\sigma$ and $\rho$ in a general and comprehensive manner, and establish precise criteria for the validity of \eqref{Claim}. We also investigate different regimes of transport and theories in which the reciprocity relation \eqref{Ohm_Reciprocity} is recovered. This may occur at low energies, when photons cease to propagate, or when another mechanism makes $G^{jj}$ and $\Pi$ indistinguishable. Both classes of transport behaviour are demonstrated with explicit examples. Namely, we first study Eq.~\eqref{Claim} in the context of the theory of electrodynamics in our universe: quantum electrodynamics or QED${}_4$. We then show examples of simpler QFTs in which the reciprocity relation \eqref{Ohm_Reciprocity} and its violation \eqref{Claim} can be understood more easily and analytically.

{\bf General considerations and QED${}_4$.---}The Kubo formula \eqref{Kubo_ext} for $\sigma(\omega)$ is usually derived in a setting with a global $U(1)$ symmetry (see Refs.~\cite{KapustaGale,leBellac:thermalFT}). The absence of dynamical photons (and of photon  propagators in a diagrammatic expansion) then effectively identifies $G^{jj} = \Pi$. Hence, $\sigma = \tilde \sigma$, and the resistivity can only be given by $\tilde\rho$ from Eq.~\eqref{Kubo_int}, which is by definition the inverse of $\tilde \sigma$. 

Gauging the global $U(1)$ symmetry then introduces one-particle-reducible (1PR) diagrams from two loops on, distinguishing $G^{jj}$ and $\Pi$. One may thus expect that, at least, $\sigma \neq \tilde \sigma$. However, as we argue below, this depends on whether the state has $T=0$ or $T\neq0$.

\begin{figure}
    \centering
    \includegraphics[width=\linewidth]{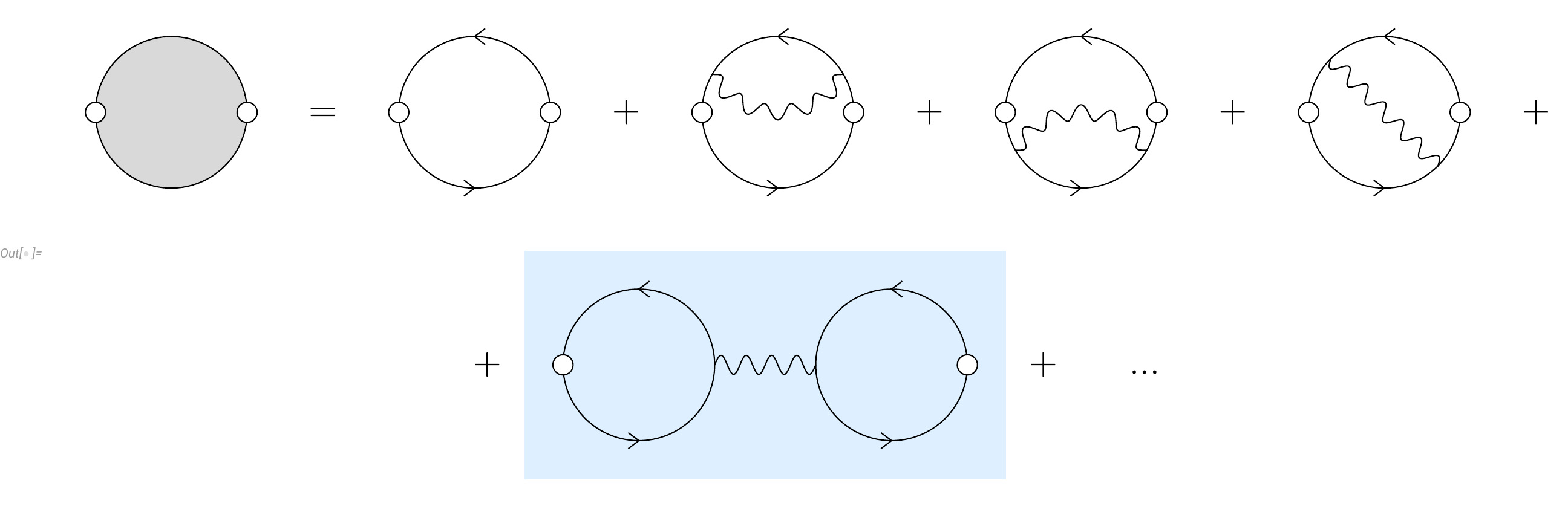}
    \caption{Diagrammatic expansion of $G^{jj}$ in QED${}_4$ to two loops. The diagrams in the first line contribute to the 1PI computation of $\Pi$. The 1PR diagram in a blue box does not.}
    \label{fig:diags}
\end{figure}

We start by considering a $T=0$ state in QED${}_4$ with a single massive Dirac fermion. The diagrams up to two loops contributing to $G^{jj}$ are depicted in Fig.~\ref{fig:diags} and we show in Appendix~\ref{app:qedT0} that, indeed, $G^{jj}\neq\Pi$. While $\sigma$ and $\rho$, as defined in \eqref{Kubo_ext}, have little physical meaning in 4$d$ at $T=0$, it is nevertheless instructive to formally compute $\sigma \rho$ and investigate Eq.~\eqref{Claim}. First, we express $G^{EE}$ in terms of $G^{AA}$ to write the Kubo formula \eqref{Kubo_ext} for $\rho$ as
\begin{equation}
    \label{eq:naive_resA}
    \rho(\omega) = \lim_{k\to 0} i\omega \left( G^{AA}_{R}(\omega,k) - \Box^{-1}(\omega,k) \right),
\end{equation}
where $\Box$ is the classical photon kinetic operator. For purposes of perturbative calculations, we can use $\Box^{-1}(\omega,0) =  G_\text{free}^{AA}(\omega,0) = 1/\omega^{2}$. We then observe that $G_F^{jj}$ is the bubble containing all loop corrections to the photon propagator, resumming diagrams beyond Fig.~\ref{fig:diags}. It can be obtained by removing its free part and amputating the remaining external legs (see Ref.~\cite{WeinbergFoundations}):
\begin{equation}
    \label{eq:imp_vac_pol}
    G^{jj}_F = \Box \cdot ( G^{AA}_F  - G^{AA}_\text{free} ) \cdot  \Box,
\end{equation}
where the arguments $(\omega,k)$ have been removed for brevity. Analytically continuing \eqref{eq:imp_vac_pol} to $G_R^{jj}$ and plugging it into \eqref{Ohm_ext} gives an alternative expression for $\sigma(\omega)$, which allows us to formally express the reciprocity relation as 
\begin{equation}
    \label{eq:non_inv_rel}
    \sigma(\omega) \rho(\omega) = \lim_{k\to 0} \left( \frac{\pi(\omega,k)}{ \pi(\omega,k) - 1} \right)^{2},
\end{equation}
where $\pi$ is the scalar dimensionless vacuum polarisation that enters $G_F^{AA}$. Eq.~\eqref{eq:non_inv_rel} therefore shows that, generically, Ohm's reciprocity relation is violated at $T=0$, confirming Eq.~\eqref{Claim}.  To one loop in $\alpha = e^2/4\pi$, $\pi(\omega,k=0) = \alpha \omega^2 / 15 \pi m^2 + O(\omega^4)$ \cite{laporta2loop} and therefore 
\begin{equation}\label{QED_res_zerotemp}
    \sigma(\omega) \rho(\omega) = \frac{\alpha^2 \omega^4}{225 \pi^2 m^4}  + O(\omega^6).
\end{equation}

\begin{figure}
    \centering
    \includegraphics[width=\linewidth]{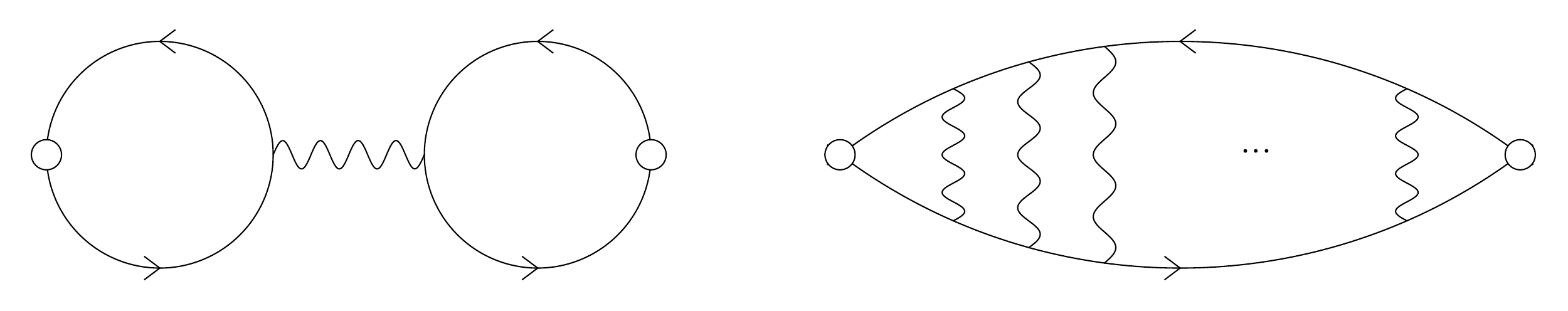}
    \caption{Two types of diagrams contributing to the leading-order perturbative expansion of $G^{jj}$ at $T \neq 0$ \cite{Jeon:long}.} 
    \label{fig:kernels}
\end{figure}

Eq.~\eqref{eq:non_inv_rel} also holds at $T\neq0$, but for finite (non-zero) $\omega$, with $\pi$ computed e.g.~in Ref.~\cite{leBellac:thermalFT}.

In the DC limit, however, the situation is dramatically different because the perturbative loop expansion of Fig.~\ref{fig:diags} breaks down. Instead, an infinite subset of higher-loop diagrams, such as the ones depicted in Fig.~\ref{fig:kernels}, may contribute at the leading order of the calculation \cite{Jeon:long,AMYpaper}. The diagrams, which can be classified based on their topology, need to be resummed to obtain correct transport coefficients. At high temperature\footnote{By `high temperature' we mean temperatures much larger than any mass scale in the theory \cite{AMYpaper}.} and in the weakly coupled regime of $\alpha$ where the leading-log approximation applies, it can be shown that the 1PR diagrams contributing to $G^{jj}$ are suppressed. This effectively identifies $\sigma = \tilde\sigma$. On the other hand, the thermal dressing of photons implies that, due to the definition of $\Pi$, $G^{AA} \simeq \Pi^{-1}$ in the DC limit. As a consequence, the thermal DC limit reinstates the reciprocity relation \eqref{Ohm_Reciprocity} in QED${}_4$ and other similar thermal field theories. This classification of diagrams and the relevant details of the approximation needed to derive Eq.~\eqref{Ohm_Reciprocity} are discussed in Appendix~\ref{app:qedT}.

It is interesting to note that, qualitatively, the same result is obtained by simply using the small-$\omega$ limit of the one-loop, hard thermal loop expression for $\pi$ given in \cite{Weldon_plasma}. Using $\pi(\omega,k=0) = 4\pi \alpha T^2 / 9 \omega^2 + O(\omega^0)$, one finds
\begin{equation}
    \sigma(\omega) \rho(\omega) = 1 + \frac{9\omega^2}{2\pi \alpha T^2} + O(\omega^4).
\end{equation}
While informative, this is a heuristic result as $\pi$ used here is computed outside the strict hydrodynamic regime.

In summary, while, in general, Ohm's reciprocity is violated (Eq.~\eqref{Claim}) for AC transport in QED${}_4$, thermal effects can reestablish \eqref{Ohm_Reciprocity} in the DC limit. Importantly, it is also clear that, generically, $\sigma \neq \tilde\sigma$, $\rho \neq \tilde\rho$, and that context needs to be provided to be able to choose the physically relevant definition. It should, however, be born in mind that the analysis presented here is inherently perturbative and its details are not expected to apply to the non-perturbative regime studied in Ref.~\cite{das2023higherform}, which found that $\sigma \rho \neq 1$ even in the DC limit of scalar QED${}_4$.

Suppression of dynamical photons in the hydrodynamic regime can occur via a number of mechanisms, either only for DC transport or even for AC transport. Below, we study two simpler classes of QFTs to further elucidate the conditions needed to reinstate Eq.~\eqref{Ohm_Reciprocity}:

$\bullet$ 3$d$ CFTs in which the kinetic photon terms are irrelevant in the IR. Such theories typically exhibit a particle-vortex duality and Ohm's reciprocity \eqref{Ohm_Reciprocity} even at $T=0$. Fascinatingly, however, special 3$d$ CFTs exist that evade these conditions and violate the reciprocity relation.

$\bullet$ Holographic supersymmetric large-$N$ QFTs with a one-loop exact electromagnetic beta function.

{\bf Three-dimensional CFTs and particle-vortex duality.---}Three-dimensional QFTs frequently possess more symmetries than those in 4$d$ and exhibit dualities, such as the prominent particle-vortex duality that was long known in bosonic theories and has recently been extended to theories with fermions \cite{MetlitskiVishwanath,Son_compositefermion,SeibergWitten_web,Karch:2016sxi}. Many 3$d$ theories also exhibit fixed points and can be understood as CFTs. In fact, it was argued in Ref.~\cite{Witten:2003ya} that a generic CFT${}_3$ with a global $U(1)$ current has a particle-vortex dual (an S-dual) theory (see also \cite{hartnoll2018holographic}). 

We begin our analysis of such theories by studying the so-called QED$_{3,4}$ (see Refs.~\cite{GorbarGusynin,Teber_2012,Son_compositefermion,Son_selfdual}), which is a theory of a Dirac fermion $\Psi$ (one can also consider a theory with $N$ flavours) confined to a 3$d$ brane that interacts with a photon propagating in 4$d$ half-space:
\begin{equation}
    \label{eq:PV_el}
    S_\Psi = \int d^3 x \, i \bar\Psi \gamma^b \left(\partial_b - i A_b\right) \Psi - \frac{1}{4e^2} \int d^4 x \,F_{\mu\nu} F^{\mu\nu}.
\end{equation}
At weak coupling $e$, the theory is scale invariant. At strong coupling, it is known to be scale invariant for large $N > N_{\rm crit}$. For small $N$, the theory may undergo spontaneous chiral symmetry breaking and generate a fermion mass \cite{GorbarGusynin}, thereby remaining scale invariant only for some $e < e(N_{\rm crit})$. Whether a genuine CFT${}_3$ for $N=1$ or not, QED${}_{3,4}$ in \eqref{eq:PV_el} is conjectured to be particle-vortex dual, at least in the IR, to a theory of a neutral composite Dirac fermion $\psi$ coupled to an emergent gauge field $a_b$: 
\begin{align}
    \label{eq:PV_cf}
    \bar S_\psi =&\, \int d^3 x \left[ i \bar\psi \gamma^b \left(\partial_b - i a_b\right) \psi - \frac{1}{4\pi} \varepsilon^{abc} A_a \partial_b a_c \right] \nn
   &- \frac{1}{4e^2} \int d^4 x \,F_{\mu\nu} F^{\mu\nu}.
\end{align}
The weak-strong duality relates the fermionic $U(1)$ currents to the topological currents (see Ref.~\cite{Son_selfdual}):
\begin{align}
    j_\Psi^\mu = \bar\Psi \gamma^\mu \Psi &\leftrightarrow - \frac{1}{4\pi} \varepsilon^{\mu\nu\lambda} \partial_\nu a_\lambda ,\\
    j_\psi^\mu = \bar\psi \gamma^\mu \psi &\leftrightarrow  \frac{1}{4\pi} \varepsilon^{\mu\nu\lambda}\partial_\nu A_\lambda ,
\end{align}
or, in terms of spatial components of currents and electric fields in two dual theories ($E^j$ is defined in terms of $A^\mu$ and $e^j$ in terms of $a^\mu$), 
\begin{equation}\label{duality_transform}
    j^i_\Psi \leftrightarrow - \frac{1}{4\pi} \varepsilon^{ij} e_j , \qquad j^i_\psi \leftrightarrow \frac{1}{4\pi} \varepsilon^{ij} E_j,
\end{equation}
with $\varepsilon^{ij}$ the 2$d$ Levi-Civita symbol. The couplings are dualised as $e \leftrightarrow \bar e = 8\pi/e$. All quantities computed in $\bar S_{\psi}$ will be given overhead bars. In each of the theories, we now assume (anisotropic) Ohm's law: $j_\Psi^i = \sigma^{ij} E_j$ and $j_\psi^i = \bar \sigma^{ij} e_j$. Using the duality \eqref{duality_transform}, it then follows that 
\begin{equation}\label{duality_relation_sigmas}
    \sigma^{il} = - \frac{1}{(4\pi)^2} \varepsilon^{ij} \bar\sigma^{-1}_{jk} \varepsilon^{kl}  .
\end{equation}
Moreover, the correlators that enter the Kubo formulae \eqref{Kubo_ext} (for any type of $G^{jj}$ and $G^{EE}$) are dualised as
\begin{align}
    \langle j_\Psi^i , j_\Psi^j \rangle &\leftrightarrow  \frac{1}{(4\pi)^2} \varepsilon^{ik} \varepsilon^{jl} \langle e_k, e_l \rangle, \\
    \langle j_\psi^i ,j_\psi^j \rangle  &\leftrightarrow  \frac{1}{(4\pi)^2} \varepsilon^{ik} \varepsilon^{jl} \langle E_k, E_l \rangle,
\end{align}
which relates $\sigma(e)$ to $\bar \rho(\bar e)$ and $\sigma(\bar e)$ to $\rho(e)$ as
\begin{align}
    \sigma^{ij}(e) &= \frac{1}{(4\pi)^2} \varepsilon^{ik} \varepsilon^{jl} \bar \rho_{kl}(\bar e), \\
    \bar \sigma^{ij}(\bar e) & = \frac{1}{(4\pi)^2} \varepsilon^{ik} \varepsilon^{jl} \rho_{kl} (e).
\end{align}
Thus, we recover Ohm's  reciprocity relations \eqref{Ohm_Reciprocity} (in matrix form) in both dual theories: $\sigma\cdot \rho = 1$ and $\bar\sigma \cdot \bar \rho = 1$, at finite temperature $T$, and for all $\omega$ and couplings $e$. 

We continue with a discussion of general 3$d$ CFTs at zero temperature. In such theories, $G^{jj}_R$ is constrained by the relativistic symmetries to have the form
\begin{equation}\label{CFT_current}
G^{j^\mu j^\nu}_R(p) = \sqrt {p^2} \pars{\eta^{\mu\nu} - \frac{p^\mu p^\nu}{p^2}} C_j,
\end{equation}
where $p^\mu$ is the relativistic three-momentum and $C_j$ is momentum-independent constant. If the CFT${}_3$ has a number of conserved (non-commuting) currents $j^\mu_I$ labeled by $I$, then we have a tensor $C^{IJ}_j$ of constants. The form \eqref{CFT_current} is sufficient to determine the conductivity at $T=0$, and, equivalently, in the $\omega \gg T$ limit of the AC conductivity in a thermal state (see Refs.~\cite{Fisher_Hall,Mtheory,hartnoll2018holographic,HuhSachdev}):
\begin{equation}\label{Sigma_Cj}
    \sigma(\omega \gg T) \approx \sigma(T=0) = C_j .
\end{equation}
The resistivity \eqref{Kubo_ext} can be computed from the photon correlator $G^{A^\mu A^\nu}_R$ (cf.~Eq.~\eqref{eq:naive_resA}) that has the same tensor structure as Eq.~\eqref{CFT_current} (in the appropriate gauge) but a generically independent constant $C_A$. We find 
\begin{equation}\label{Rho_CA}
    \rho(\omega \gg T) \approx \rho(T=0) = C_A .
\end{equation}
Evaluation of the violation of Ohm's reciprocity \eqref{Claim} is therefore reduced to computing $C_j$ and $C_A$, which can be done diagrammatically. This is, unsurprisingly, simplified in CFTs with a large number of flavours $N$. In the rest of this section, we will focus only on such theories. 

We first consider QED${}_3$ with $N$ flavours of fermions $\psi_I$ at $T=0$ and in the $1/N$ expansion (see Ref.~\cite{Romatschke:2019qbx}):
\begin{equation}\label{QED3}
    S = \int d^3 x \left[ i \sum_{I=1}^N \bar\psi_I \gamma^\mu \left(\partial_\mu - i A_\mu \right) \psi_I - \frac{1}{4e^2} F_{\mu\nu} F^{\mu\nu} \right] .
\end{equation}
In this theory, the gauged `electric' current that couples to $A_\mu$ is $j^\mu = \sum_{I = 1}^N \bar \psi_I \gamma^\mu \psi_I$, which is a commuting `Casimir operator'. For a detailed analysis of all relevant aspects of this theory, see Ref.~\cite{GiombiKlebanov}. Importantly, after one integrates out the fermions, it can be seen that the photon propagator is an irrelevant operator, so that $G^{AA} = \Pi$. Moreover, in this theory, $A^\mu$ and $j^\mu$ are related by a Legendre transform \cite{Witten:2003ya,Leigh:2003ez}. The computation of $C_j$ and $C_A$ was also performed in \cite{GiombiKlebanov}, giving
\begin{align}
        \sigma = C_j &= \frac{N}{16} \left[ 1 + \frac{1}{N}\pars{\frac{736}{9\pi^2}- 8} + O(1/N^2)\right],\\
        \rho = C_A &= \frac{16}{N} \left[ 1 - \frac{1}{N}\pars{\frac{736}{9\pi^2}- 8} + O(1/N^2) \right].
\end{align}
Hence, as expected in such 3$d$ CFTs, $\sigma \rho = 1$.

Finally, we turn our attention to a very special but instructive CFT${}_3$: the $\mathbb{C}P^{N-1}$ model with the action
\begin{equation}\label{CP}
    S = \frac{1}{e^2} \int d^3 x \left|\left( \partial_\mu - i A_\mu \right) \varphi \right|^2 ,
\end{equation}
where the $N$ complex scalars $\varphi_I$ satisfy the constraint $|\varphi|^2 = \sum_{I=1}^N |\varphi_I|^2 = 1$. For details, see Refs.~\cite{Coleman_Aspects,Polyakov:GFaS}. Remarkably, its unique properties allow us to evade the above-discussed, `generally expected' properties of a 3$d$ CFT with dynamical electromagnetism. The reason is that the action \eqref{CP} has {\em no} kinetic term for the photon. Nevertheless, such a term is dynamically generated by quantum corrections in the $1/N$ expansion. Crucially, it is now not irrelevant. The QFT therefore has dynamical photons in the IR that gauge the current $j^\mu = - i \sum_{I=1}^N \left[\varphi_I^\dagger D_{\mu} \varphi_I - \varphi_I (D_{\mu} \varphi^I)^\dagger \right]$ (see Ref.~\cite{HuhSachdev}). The theory at $T=0$ offers an explicit example of the Ohm's reciprocity relation violation, i.e., of Eq.~\eqref{Claim}. This again follows from the coefficients $C_j$ and $C_A$ that were computed in Ref.~\cite{HuhSachdev}, giving us
\begin{align}
     \sigma= C_j &= \frac{N}{16} \left[ 1 - \frac{2.74}{N} + O(1/N^2) \right], \\
     \rho = C_A &= \frac{16}{N} \left[ 1 + \frac{0.578}{N} + O(1/N^2) \right].
\end{align}
The leading term still preserves $\sigma \rho = 1$. However, the $1/N$ corrections, which introduce dynamical photons, violate the reciprocity relation: 
\begin{equation}
    \sigma \rho = 1 - \frac{2.16}{N} + O(1/N^2).
\end{equation}
As in QED${}_4$, this example elucidates the essential role that dynamical IR photons play in the violation of Ohm's reciprocity relation \eqref{Claim}. It is also important to note that since this is a $T=0$ calculation, no screening protects Ohm's reciprocity even for DC transport coefficients. 

{\bf Supersymmetric field theories and holographic duality.---}Holographic duality is a powerful tool for computing correlation functions in certain large-$N$ QFTs \cite{zaanen2015holographic,hartnoll2018holographic}.  Consider the $\CN = 4$ supersymmetric Yang-Mills theory with $N \to \infty$ number of colours and an infinite 't Hooft coupling. We are interested in a scenario in which a current $j^\mu$ that corresponds to a $U(1)$ subgroup of the R-symmetry is gauged so that the QFT has dynamical electromagnetism with a one-loop exact beta function of the running electromagnetic coupling and a UV Landau pole. For details, see Refs.~\cite{Fuini:2015hba,Grozdanov:2017kyl,Hofman:2017vwr}. Correlators of ${\bf j}$ (giving $\sigma$) and ${\bf E}$ (giving $\rho$) can be computed by using two different 5$d$ bulk actions with mixed boundary conditions:
\begin{align}
    S_\sigma &= \int d^5 x \sqrt{-g} \spars{ R - 2 \Lambda - \frac{1}{4e^2} F_{\mu\nu} F^{\mu\nu} },  \label{eq:EM_action} \\
    S_\rho &= \int d^5 x \sqrt{-g} \spars{ R - 2 \Lambda - \frac{1}{12e^2} H_{\mu\nu\rho} H^{\mu\nu\rho} } \label{eq:HF_action},
\end{align}
where $\Lambda = - 6 / L^2$. $L$ is the AdS scale, which we set to one. Eq.~\eqref{eq:EM_action} is the 5$d$ Einstein-Maxwell theory and Eq.~\eqref{eq:HF_action} the theory of a two-form field $B_{\mu\nu}$ with a three-form field strength $H=dB$ developed in \cite{Grozdanov:2017kyl,Hofman:2017vwr} (see also \cite{Grozdanov:2018fic,DeWolfe:2020uzb}) in constructing a holographic dual to a theory with a one-form symmetry and magnetohydrodynamics with dynamical electromagnetism. The actions \eqref{eq:EM_action} and \eqref{eq:HF_action} are related by the 5$d$ Hodge dualisation $H=*_5 F$.

We assume electric charge neutrality and neglect the energy-momentum fluctuations (the probe limit). 
The relevant fluctuations of the bulk fields in the background of the Schwarzschild black brane in the two cases are $\delta A = \delta A (u) \, e^{-i \omega t} dx$ and $\delta B = \delta B (u) \, e^{-i \omega t} dy \wedge dz$, where $u$ is the radial Fefferman-Graham (FG) coordinate  (see Refs.~\cite{Skenderis_holo_renormalization,Taylor_counterterms}). Following the procedure of Ref.~\cite{Grozdanov:2017kyl} to evaluate the Kubo formulae \eqref{Kubo_ext} then gives
\begin{align}
    \sigma(\omega) &= \frac{2 \delta A^{(2)}}{i\omega \delta A^{(0)}} + i \omega \pars{\ln \bar u + \frac{1}{2}},\label{eq:hol_sigma} \\
    \rho (\omega) &= \frac{\delta B^{(1)}}{i \omega \pars{\delta B^{(0)} + \delta B^{(1)} \ln \bar u}},\label{eq:hol_rho}
\end{align}
where $\delta A^{(i)}$ and $\delta B^{(i)}$ are the $i$-th order coefficients in the FG expansion, and $\bar u$ is the radial position of a finite cutoff brane introduced to regularise the theory. As a result of the dualisation $H=*_5 F$, we have $\omega \delta A^{(0)} = i \delta B^{(1)}$ and $2 \delta A^{(2)} = i \omega ( \delta B^{(0)} - \delta B^{(1)}/2)$. Inserting these relations into Eqs.~\eqref{eq:hol_sigma} and \eqref{eq:hol_rho} then immediately establishes that Ohm's reciprocity relation \eqref{Ohm_Reciprocity} is retained for the AC transport coefficients: $\sigma(\omega)\rho(\omega) = 1$.
In fact, Eq.~\eqref{Ohm_Reciprocity} holds in a large class of holographic models, discussed in detail in Ref.~\cite{Giorgio}. The large-$N$ and supersymmetric nature of this holographic QFT therefore ensures that Ohm's reciprocity \eqref{Ohm_Reciprocity} is preserved for all frequencies.

{\bf Discussion.---}Our work shows that one should approach conventional wisdom with care. In particular, when it comes to the measurement of electric response, one needs to carefully and precisely specify the nature of `external' perturbations, as per Eqs.~\eqref{Ohm_ext} and \eqref{Ohm_int}. When this is done in the usual sense of Eq.~\eqref{Ohm_ext}, then one arrives at the conclusion that Ohm's reciprocity is generically violated (Eq.~\eqref{Claim}), except in special circumstances when the dynamics of electromagnetism is effectively suppressed, most notably in the perturbative DC limit with thermal effects. Phenomenologically, Eq.~\eqref{Claim} is important as it refers to transport given in terms of the IR conserved operators that dominate late-time dynamics, and not in terms of the 1PI vacuum polarisation of the photons. We believe it should be possible to devise simpler (e.g., non-relativistic lattice) models to show this behaviour and potentially even measure the differences between AC $\sigma$ and $1/\rho$ in experiments. Theories that will exhibit DC violation seem harder to access in real world at least in states with hydrodynamic behaviour. This is because the DC relation $\sigma = 1/\rho$ appears naturally in the derivation of magnetohydrodynamics \cite{Grozdanov:2016tdf,Hernandez:2017mch,Armas:2018atq}. Finally, to develop better understanding of the differences between conductive and resistive transport in QFTs beyond linear response theory, it should be most transparent to further develop and employ the Schwinger-Keldysh effective field theory methods \cite{Polonyi:2014rpa,Polonyi:2015cna,Glorioso:2018wxw}, already used in QED${}_4$ in Ref.~\cite{Grozdanov:2015nea}.

{\bf Acknowledgements.---} We would like to thank Arpit Das, Sean Hartnoll, Nabil Iqbal, Janos Polonyi, Nick Poovuttikul, Toma\v{z} Prosen, Paul Romatschke, Alexander Soloviev and Mile Vrbica for illuminating discussions on related topics. The work of G.F. is supported by an Edinburgh Doctoral College Scholarship (ECDS). The work of S.G. was supported by the STFC Ernest Rutherford Fellowship ST/T00388X/1. The work is also supported by the research programme P1-0402 and the project N1-0245 of Slovenian Research Agency (ARIS).

\appendix
\setcounter{secnumdepth}{2}
\section{QED${}_4$ at zero temperature}
\label{app:qedT0}

In this appendix, we briefly review some relevant results in QED${}_4$ computed to two loops at $T=0$ (see Ref.~\cite{laporta2loop}). In particular, the current correlator $G^{jj}_F$ is given in the low frequency limit and at $k=0$ by 
\begin{align}
        \label{eq:2l_cond_jj}
        &G_F^{jj}(\omega) = \omega^2 \bigg[ \bigg(\frac{\alpha}{15\pi} + \frac{41 \alpha^2}{162 \pi^2} \bigg) \frac{\omega^2}{m^2} + \nn 
         &+ \bigg(\frac{\alpha}{140\pi} + \frac{497 \alpha^2}{10800 \pi^2} \bigg) \frac{\omega^4}{m^4} \bigg] + O \left(\alpha^3, \frac{\omega^6}{m^6} \right),
\end{align}
while the 1PI vacuum polarisation is given by 
\begin{align}
        \label{eq:2l_cond_pi}
        &\Pi(\omega) = \omega^2 \bigg[ \bigg(\frac{\alpha}{15\pi} + \frac{41 \alpha^2}{162 \pi^2} \bigg) \frac{\omega^2}{m^2} + \nn 
        &+ \bigg(\frac{\alpha}{140\pi} + \frac{449 \alpha^2}{10800 \pi^2} \bigg) \frac{\omega^4}{m^4} \bigg] + O \bigg( \alpha^3, \frac{\omega^6}{m^6} \bigg).
\end{align}

The difference between Eqs.~\eqref{eq:2l_cond_jj} and \eqref{eq:2l_cond_pi} can be traced back precisely to the 1PR diagram (shaded in blue) in Fig.~\ref{fig:diags}. These results are sufficient to formally compute $\sigma$, $\tilde\sigma$, $\rho$, $\tilde\rho$, and, thereby, show that $\sigma\rho \neq 1$, as stated in Eq.~\eqref{QED_res_zerotemp}.

\section{Finite temperature field theories and the breakdown of perturbation theory}
\label{app:qedT}

Here, we show in more detail how the DC ($\omega\to0$) limit in a thermal field theory causes the breakdown of the perturbative diagrammatic expansion used in Fig.~\ref{fig:diags}. This necessitates a resummation of higher-loop diagrams to extract the correct hydrodynamic behaviour of thermal correlators. We then use this fact to argue that, at the leading order in the coupling and in the DC limit, the reorganised expansion leads to the conclusion that $\sigma = \tilde\sigma$ and $\sigma = 1/\rho$. I.e., Ohm's reciprocity relation \eqref{Ohm_Reciprocity} is recovered. The arguments presented here rely on the thermal field theory and kinetic theory results presented in Refs.~\cite{Jeon:long,AMYpaper}.

The crux of the argument is that in calculating thermal correlators, one encounters loop integrals containing a product of two renormalised Euclidean propagators (we suppress spatial momentum dependence): $\sum\nolimits_{\omega_n} \tilde G(\omega) \tilde G(\omega+\delta)$, where $\delta$ is an external frequency, and $\omega_n$ are the appropriate Matsubara frequencies. Because of interactions and finite temperature, each propagator has 4 simple poles at $\pm E_p \pm i \Gamma_p$ in the complex frequency plane, where the width $\Gamma_p$ is nonzero because the vacuum polarisation develops an imaginary part when $T\neq0$. For finite $\delta$, this does not produce any notable issues. However, in the coincident $\delta\to0$ limit -- the DC limit --, all poles becomes of second order and their contribution to the loop summation is of the order $O(1/\Gamma_p) = O(1/e^2)$. As a consequence, a diagram with $n$ explicit vertices and $m$ pairs of propagators with equal momenta, and with any number of loops, may be enhanced up to order $O(e^{n-2m})$. The fact that new types of diagrams with different numbers of vertices and propagators can scale the same way with the coupling $e \ll 1$ requires us to reorganise the usual perturbative series.

Following Ref.~\cite{Jeon:long}, these diagrams can be grouped on the basis of how many equal-momentum propagators they may contain. The diagram (topologies) that, at least naively, contribute at the same, leading order as the simple loop diagram are the ones depicted in the right panel of of Fig.~\ref{fig:kernels}. We refer to the type of diagram in the left panel as the chain diagram and to the relevant ones on the right as ladder diagrams. Ladder diagrams in which the rungs are fermions rather than photons are also possible, an example being the diagram discussed in Ref.~\cite{Grozdanov:2016tdf}. While diagrams with any number of rungs exist, there is only one chain diagram. This is a consequence of each line representing a dressed propagator.

To identify the diagrams that capture the relevant DC behaviour, one must analyse all possible processes that map two intermediate particles into another two. For fermions in single-flavour QED${}_4$, this translates to finding out which, if any, momentum-transfer channel ($s$, $t$ or $u$) dominates over the others in a $2\to2$ scattering process. This is a difficult question that does not have a general answer valid over all possible temperature or coupling regimes. Ref.~\cite{AMYpaper}, however, derived a result valid at high temperatures and at the leading logarithm order of the coupling. There, it was shown that the $s$-channel exchanges (corresponding to the chain diagram in Fig.~\ref{fig:kernels}) are in effect subleading with respect to the $t$- and $u$-channel processes. Because of that, we conclude that, within the regime of validity of the assumptions of the original work, the 1PR part of $G^{jj}$ is effectively suppressed. Hence, in this limit, we find that $\sigma = \tilde\sigma$ and that the Ohm's reciprocity relation $\sigma = 1/\rho$ (cf.~Eq.~\eqref{Ohm_Reciprocity}) is recovered. Extending these results outside of this regime is a rather technical task which would bring us far away from the scope of this work, and we thus leave it to the future.

\bibliography{lit.bib}

\begin{thebibliography}{46}%
\makeatletter
\providecommand \@ifxundefined [1]{%
 \@ifx{#1\undefined}
}%
\providecommand \@ifnum [1]{%
 \ifnum #1\expandafter \@firstoftwo
 \else \expandafter \@secondoftwo
 \fi
}%
\providecommand \@ifx [1]{%
 \ifx #1\expandafter \@firstoftwo
 \else \expandafter \@secondoftwo
 \fi
}%
\providecommand \natexlab [1]{#1}%
\providecommand \enquote  [1]{``#1''}%
\providecommand \bibnamefont  [1]{#1}%
\providecommand \bibfnamefont [1]{#1}%
\providecommand \citenamefont [1]{#1}%
\providecommand \href@noop [0]{\@secondoftwo}%
\providecommand \href [0]{\begingroup \@sanitize@url \@href}%
\providecommand \@href[1]{\@@startlink{#1}\@@href}%
\providecommand \@@href[1]{\endgroup#1\@@endlink}%
\providecommand \@sanitize@url [0]{\catcode `\\12\catcode `\$12\catcode
  `\&12\catcode `\#12\catcode `\^12\catcode `\_12\catcode `\%12\relax}%
\providecommand \@@startlink[1]{}%
\providecommand \@@endlink[0]{}%
\providecommand \url  [0]{\begingroup\@sanitize@url \@url }%
\providecommand \@url [1]{\endgroup\@href {#1}{\urlprefix }}%
\providecommand \urlprefix  [0]{URL }%
\providecommand \Eprint [0]{\href }%
\providecommand \doibase [0]{https://doi.org/}%
\providecommand \selectlanguage [0]{\@gobble}%
\providecommand \bibinfo  [0]{\@secondoftwo}%
\providecommand \bibfield  [0]{\@secondoftwo}%
\providecommand \translation [1]{[#1]}%
\providecommand \BibitemOpen [0]{}%
\providecommand \bibitemStop [0]{}%
\providecommand \bibitemNoStop [0]{.\EOS\space}%
\providecommand \EOS [0]{\spacefactor3000\relax}%
\providecommand \BibitemShut  [1]{\csname bibitem#1\endcsname}%
\let\auto@bib@innerbib\@empty
\bibitem [{\citenamefont {Kittel}(2004)}]{Kittel}%
  \BibitemOpen
  \bibfield  {author} {\bibinfo {author} {\bibfnamefont {C.}~\bibnamefont
  {Kittel}},\ }\href@noop {} {\emph {\bibinfo {title} {Introduction to Solid
  State Physics}}}\ (\bibinfo  {publisher} {Wiley},\ \bibinfo {address} {New
  York},\ \bibinfo {year} {2004})\BibitemShut {NoStop}%
\bibitem [{\citenamefont {Mahan}(2000)}]{mahan}%
  \BibitemOpen
  \bibfield  {author} {\bibinfo {author} {\bibfnamefont {G.~D.}\ \bibnamefont
  {Mahan}},\ }\href@noop {} {\emph {\bibinfo {title} {Many Particle Physics,
  Third Edition}}}\ (\bibinfo  {publisher} {Plenum},\ \bibinfo {address} {New
  York},\ \bibinfo {year} {2000})\BibitemShut {NoStop}%
\bibitem [{\citenamefont {Kovtun}(2012)}]{Kovtun:2012rj}%
  \BibitemOpen
  \bibfield  {author} {\bibinfo {author} {\bibfnamefont {P.}~\bibnamefont
  {Kovtun}},\ }\bibfield  {title} {\bibinfo {title} {{Lectures on hydrodynamic
  fluctuations in relativistic theories}},\ }\href
  {https://doi.org/10.1088/1751-8113/45/47/473001} {\bibfield  {journal}
  {\bibinfo  {journal} {J. Phys. A}\ }\textbf {\bibinfo {volume} {45}},\
  \bibinfo {pages} {473001} (\bibinfo {year} {2012})}\BibitemShut {NoStop}%
\bibitem [{\citenamefont {Hartnoll}\ \emph {et~al.}(2018)\citenamefont
  {Hartnoll}, \citenamefont {Lucas},\ and\ \citenamefont
  {Sachdev}}]{hartnoll2018holographic}%
  \BibitemOpen
  \bibfield  {author} {\bibinfo {author} {\bibfnamefont {S.}~\bibnamefont
  {Hartnoll}}, \bibinfo {author} {\bibfnamefont {A.}~\bibnamefont {Lucas}},\
  and\ \bibinfo {author} {\bibfnamefont {S.}~\bibnamefont {Sachdev}},\
  }\href@noop {} {\emph {\bibinfo {title} {Holographic Quantum Matter}}}\
  (\bibinfo  {publisher} {MIT Press},\ \bibinfo {address} {Cambridge},\
  \bibinfo {year} {2018})\BibitemShut {NoStop}%
\bibitem [{\citenamefont {Zaanen}\ \emph {et~al.}(2015)\citenamefont {Zaanen},
  \citenamefont {Liu}, \citenamefont {Sun},\ and\ \citenamefont
  {Schalm}}]{zaanen2015holographic}%
  \BibitemOpen
  \bibfield  {author} {\bibinfo {author} {\bibfnamefont {J.}~\bibnamefont
  {Zaanen}}, \bibinfo {author} {\bibfnamefont {Y.}~\bibnamefont {Liu}},
  \bibinfo {author} {\bibfnamefont {Y.}~\bibnamefont {Sun}},\ and\ \bibinfo
  {author} {\bibfnamefont {K.}~\bibnamefont {Schalm}},\ }\href@noop {} {\emph
  {\bibinfo {title} {Holographic Duality in Condensed Matter Physics}}}\
  (\bibinfo  {publisher} {Cambridge University Press},\ \bibinfo {address}
  {Cambridge},\ \bibinfo {year} {2015})\BibitemShut {NoStop}%
\bibitem [{\citenamefont {Kapusta}\ and\ \citenamefont
  {Gale}(2011)}]{KapustaGale}%
  \BibitemOpen
  \bibfield  {author} {\bibinfo {author} {\bibfnamefont {J.~I.}\ \bibnamefont
  {Kapusta}}\ and\ \bibinfo {author} {\bibfnamefont {C.}~\bibnamefont {Gale}},\
  }\href@noop {} {\emph {\bibinfo {title} {{Finite-temperature field theory:
  Principles and applications}}}},\ Cambridge Monographs on Mathematical
  Physics\ (\bibinfo  {publisher} {Cambridge University Press},\ \bibinfo
  {address} {Cambridge},\ \bibinfo {year} {2011})\BibitemShut {NoStop}%
\bibitem [{\citenamefont {Bellac}(2011)}]{leBellac:thermalFT}%
  \BibitemOpen
  \bibfield  {author} {\bibinfo {author} {\bibfnamefont {M.~L.}\ \bibnamefont
  {Bellac}},\ }\href@noop {} {\emph {\bibinfo {title} {{Thermal Field
  Theory}}}},\ Cambridge Monographs on Mathematical Physics\ (\bibinfo
  {publisher} {Cambridge University Press},\ \bibinfo {address} {Cambridge},\
  \bibinfo {year} {2011})\BibitemShut {NoStop}%
\bibitem [{\citenamefont {Chaikin}\ and\ \citenamefont
  {Lubensky}(1995)}]{chaikin_lubensky}%
  \BibitemOpen
  \bibfield  {author} {\bibinfo {author} {\bibfnamefont {P.~M.}\ \bibnamefont
  {Chaikin}}\ and\ \bibinfo {author} {\bibfnamefont {T.~C.}\ \bibnamefont
  {Lubensky}},\ }\href@noop {} {\emph {\bibinfo {title} {Principles of
  Condensed Matter Physics}}}\ (\bibinfo  {publisher} {Cambridge University
  Press},\ \bibinfo {address} {Cambridge},\ \bibinfo {year} {1995})\BibitemShut
  {NoStop}%
\bibitem [{\citenamefont {Grozdanov}\ \emph {et~al.}(2017)\citenamefont
  {Grozdanov}, \citenamefont {Hofman},\ and\ \citenamefont
  {Iqbal}}]{Grozdanov:2016tdf}%
  \BibitemOpen
  \bibfield  {author} {\bibinfo {author} {\bibfnamefont {S.}~\bibnamefont
  {Grozdanov}}, \bibinfo {author} {\bibfnamefont {D.~M.}\ \bibnamefont
  {Hofman}},\ and\ \bibinfo {author} {\bibfnamefont {N.}~\bibnamefont
  {Iqbal}},\ }\bibfield  {title} {\bibinfo {title} {{Generalized global
  symmetries and dissipative magnetohydrodynamics}},\ }\href
  {https://doi.org/10.1103/PhysRevD.95.096003} {\bibfield  {journal} {\bibinfo
  {journal} {Phys. Rev. D}\ }\textbf {\bibinfo {volume} {95}},\ \bibinfo
  {pages} {096003} (\bibinfo {year} {2017})}\BibitemShut {NoStop}%
\bibitem [{\citenamefont {Hernandez}\ and\ \citenamefont
  {Kovtun}()}]{Hernandez:2017mch}%
  \BibitemOpen
  \bibfield  {author} {\bibinfo {author} {\bibfnamefont {J.}~\bibnamefont
  {Hernandez}}\ and\ \bibinfo {author} {\bibfnamefont {P.}~\bibnamefont
  {Kovtun}},\ }\bibfield  {title} {\bibinfo {title} {{Relativistic
  magnetohydrodynamics}},\ }\href {https://doi.org/10.1007/JHEP05(2017)001}
  {\bibfield  {journal} {\bibinfo  {journal} {JHEP}\ }\textbf {\bibinfo
  {volume} {05}}\bibinfo  {number} { (2017)},\ \bibinfo {pages}
  {001}}\BibitemShut {NoStop}%
\bibitem [{\citenamefont {Armas}\ and\ \citenamefont
  {Jain}(2019)}]{Armas:2018atq}%
  \BibitemOpen
\bibfield  {number} {  }\bibfield  {author} {\bibinfo {author} {\bibfnamefont
  {J.}~\bibnamefont {Armas}}\ and\ \bibinfo {author} {\bibfnamefont
  {A.}~\bibnamefont {Jain}},\ }\bibfield  {title} {\bibinfo {title}
  {{Magnetohydrodynamics as superfluidity}},\ }\href
  {https://doi.org/10.1103/PhysRevLett.122.141603} {\bibfield  {journal}
  {\bibinfo  {journal} {Phys. Rev. Lett.}\ }\textbf {\bibinfo {volume} {122}},\
  \bibinfo {pages} {141603} (\bibinfo {year} {2019})}\BibitemShut {NoStop}%
\bibitem [{\citenamefont {Vardhan}\ \emph {et~al.}()\citenamefont {Vardhan},
  \citenamefont {Grozdanov}, \citenamefont {Leutheusser},\ and\ \citenamefont
  {Liu}}]{Vardhan:2022wxz}%
  \BibitemOpen
  \bibfield  {author} {\bibinfo {author} {\bibfnamefont {S.}~\bibnamefont
  {Vardhan}}, \bibinfo {author} {\bibfnamefont {S.}~\bibnamefont {Grozdanov}},
  \bibinfo {author} {\bibfnamefont {S.}~\bibnamefont {Leutheusser}},\ and\
  \bibinfo {author} {\bibfnamefont {H.}~\bibnamefont {Liu}},\ }\href@noop {}
  {\bibinfo {title} {{A new formulation of strong-field magnetohydrodynamics
  for neutron stars}}},\ \Eprint {https://arxiv.org/abs/2207.01636}
  {arXiv:2207.01636 [astro-ph.HE]} \BibitemShut {NoStop}%
\bibitem [{\citenamefont {Weinberg}(2005)}]{WeinbergFoundations}%
  \BibitemOpen
  \bibfield  {author} {\bibinfo {author} {\bibfnamefont {S.}~\bibnamefont
  {Weinberg}},\ }\href@noop {} {\emph {\bibinfo {title} {{The Quantum theory of
  fields. Vol. 1: Foundations}}}}\ (\bibinfo  {publisher} {Cambridge University
  Press},\ \bibinfo {address} {Cambridge},\ \bibinfo {year} {2005})\BibitemShut
  {NoStop}%
\bibitem [{\citenamefont {Laporta}\ and\ \citenamefont
  {Jentschura}(2024)}]{laporta2loop}%
  \BibitemOpen
  \bibfield  {author} {\bibinfo {author} {\bibfnamefont {S.}~\bibnamefont
  {Laporta}}\ and\ \bibinfo {author} {\bibfnamefont {U.~D.}\ \bibnamefont
  {Jentschura}},\ }\bibfield  {title} {\bibinfo {title} {Dimensional
  regularization and two-loop vacuum polarization operator: Master integrals,
  analytic results, and energy shifts},\ }\href
  {http://dx.doi.org/10.1103/PhysRevD.109.096020} {\bibfield  {journal}
  {\bibinfo  {journal} {Physical Review D}\ }\textbf {\bibinfo {volume}
  {109}},\ \bibinfo {pages} {096020} (\bibinfo {year} {2024})}\BibitemShut
  {NoStop}%
\bibitem [{\citenamefont {Jeon}(1995)}]{Jeon:long}%
  \BibitemOpen
  \bibfield  {author} {\bibinfo {author} {\bibfnamefont {S.}~\bibnamefont
  {Jeon}},\ }\bibfield  {title} {\bibinfo {title} {{Hydrodynamic transport
  coefficients in relativistic scalar field theory}},\ }\href
  {https://doi.org/10.1103/PhysRevD.52.3591} {\bibfield  {journal} {\bibinfo
  {journal} {Phys. Rev. D}\ }\textbf {\bibinfo {volume} {52}},\ \bibinfo
  {pages} {3591} (\bibinfo {year} {1995})}\BibitemShut {NoStop}%
\bibitem [{\citenamefont {Arnold}\ \emph {et~al.}(2000)\citenamefont {Arnold},
  \citenamefont {Moore},\ and\ \citenamefont {Yaffe}}]{AMYpaper}%
  \BibitemOpen
  \bibfield  {author} {\bibinfo {author} {\bibfnamefont {P.}~\bibnamefont
  {Arnold}}, \bibinfo {author} {\bibfnamefont {G.~D.}\ \bibnamefont {Moore}},\
  and\ \bibinfo {author} {\bibfnamefont {L.~G.}\ \bibnamefont {Yaffe}},\
  }\bibfield  {title} {\bibinfo {title} {{Transport coefficients in high
  temperature gauge theories (I): leading-log results}},\ }\href
  {https://doi.org/10.1088/1126-6708/2000/11/001} {\bibfield  {journal}
  {\bibinfo  {journal} {JHEP}\ }\textbf {\bibinfo {volume} {11}}\bibinfo
  {number} { (2000)},\ \bibinfo {pages} {001}}\BibitemShut {NoStop}%
\bibitem [{\citenamefont {Weldon}(1982)}]{Weldon_plasma}%
  \BibitemOpen
\bibfield  {number} {  }\bibfield  {author} {\bibinfo {author} {\bibfnamefont
  {H.~A.}\ \bibnamefont {Weldon}},\ }\bibfield  {title} {\bibinfo {title}
  {Covariant calculations at finite temperature: The relativistic plasma},\
  }\href {https://doi.org/10.1103/PhysRevD.26.1394} {\bibfield  {journal}
  {\bibinfo  {journal} {Phys. Rev. D}\ }\textbf {\bibinfo {volume} {26}},\
  \bibinfo {pages} {1394} (\bibinfo {year} {1982})}\BibitemShut {NoStop}%
\bibitem [{\citenamefont {Das}\ \emph {et~al.}()\citenamefont {Das},
  \citenamefont {Florio}, \citenamefont {Iqbal},\ and\ \citenamefont
  {Poovuttikul}}]{das2023higherform}%
  \BibitemOpen
  \bibfield  {author} {\bibinfo {author} {\bibfnamefont {A.}~\bibnamefont
  {Das}}, \bibinfo {author} {\bibfnamefont {A.}~\bibnamefont {Florio}},
  \bibinfo {author} {\bibfnamefont {N.}~\bibnamefont {Iqbal}},\ and\ \bibinfo
  {author} {\bibfnamefont {N.}~\bibnamefont {Poovuttikul}},\ }\href@noop {}
  {\bibinfo {title} {Higher-form symmetry and chiral transport in real-time
  lattice $u(1)$ gauge theory}},\ \Eprint {https://arxiv.org/abs/2309.14438}
  {arXiv:2309.14438 [hep-th]} \BibitemShut {NoStop}%
\bibitem [{\citenamefont {Metlitski}\ and\ \citenamefont
  {Vishwanath}(2016)}]{MetlitskiVishwanath}%
  \BibitemOpen
  \bibfield  {author} {\bibinfo {author} {\bibfnamefont {M.~A.}\ \bibnamefont
  {Metlitski}}\ and\ \bibinfo {author} {\bibfnamefont {A.}~\bibnamefont
  {Vishwanath}},\ }\bibfield  {title} {\bibinfo {title} {{Particle-vortex
  duality of two-dimensional Dirac fermion from electric-magnetic duality of
  three-dimensional topological insulators}},\ }\href
  {https://doi.org/10.1103/PhysRevB.93.245151} {\bibfield  {journal} {\bibinfo
  {journal} {Phys. Rev. B}\ }\textbf {\bibinfo {volume} {93}},\ \bibinfo
  {pages} {245151} (\bibinfo {year} {2016})}\BibitemShut {NoStop}%
\bibitem [{\citenamefont {Son}(2015)}]{Son_compositefermion}%
  \BibitemOpen
  \bibfield  {author} {\bibinfo {author} {\bibfnamefont {D.~T.}\ \bibnamefont
  {Son}},\ }\bibfield  {title} {\bibinfo {title} {{Is the Composite Fermion a
  Dirac Particle?}},\ }\href {https://doi.org/10.1103/PhysRevX.5.031027}
  {\bibfield  {journal} {\bibinfo  {journal} {Phys. Rev. X}\ }\textbf {\bibinfo
  {volume} {5}},\ \bibinfo {pages} {031027} (\bibinfo {year}
  {2015})}\BibitemShut {NoStop}%
\bibitem [{\citenamefont {Seiberg}\ \emph {et~al.}(2016)\citenamefont
  {Seiberg}, \citenamefont {Senthil}, \citenamefont {Wang},\ and\ \citenamefont
  {Witten}}]{SeibergWitten_web}%
  \BibitemOpen
  \bibfield  {author} {\bibinfo {author} {\bibfnamefont {N.}~\bibnamefont
  {Seiberg}}, \bibinfo {author} {\bibfnamefont {T.}~\bibnamefont {Senthil}},
  \bibinfo {author} {\bibfnamefont {C.}~\bibnamefont {Wang}},\ and\ \bibinfo
  {author} {\bibfnamefont {E.}~\bibnamefont {Witten}},\ }\bibfield  {title}
  {\bibinfo {title} {{A Duality Web in 2+1 Dimensions and Condensed Matter
  Physics}},\ }\href {https://doi.org/10.1016/j.aop.2016.08.007} {\bibfield
  {journal} {\bibinfo  {journal} {Annals Phys.}\ }\textbf {\bibinfo {volume}
  {374}},\ \bibinfo {pages} {395} (\bibinfo {year} {2016})}\BibitemShut
  {NoStop}%
\bibitem [{\citenamefont {Karch}\ and\ \citenamefont
  {Tong}(2016)}]{Karch:2016sxi}%
  \BibitemOpen
  \bibfield  {author} {\bibinfo {author} {\bibfnamefont {A.}~\bibnamefont
  {Karch}}\ and\ \bibinfo {author} {\bibfnamefont {D.}~\bibnamefont {Tong}},\
  }\bibfield  {title} {\bibinfo {title} {{Particle-Vortex Duality from 3d
  Bosonization}},\ }\href {https://doi.org/10.1103/PhysRevX.6.031043}
  {\bibfield  {journal} {\bibinfo  {journal} {Phys. Rev. X}\ }\textbf {\bibinfo
  {volume} {6}},\ \bibinfo {pages} {031043} (\bibinfo {year}
  {2016})}\BibitemShut {NoStop}%
\bibitem [{\citenamefont {Witten}(2003)}]{Witten:2003ya}%
  \BibitemOpen
  \bibfield  {author} {\bibinfo {author} {\bibfnamefont {E.}~\bibnamefont
  {Witten}},\ }\bibfield  {title} {\bibinfo {title} {{SL(2,Z) action on
  three-dimensional conformal field theories with Abelian symmetry}},\ }in\
  \href@noop {} {\emph {\bibinfo {booktitle} {{From Fields to Strings:
  Circumnavigating Theoretical Physics: A Conference in Tribute to Ian
  Kogan}}}}\ (\bibinfo {year} {2003})\ pp.\ \bibinfo {pages}
  {1173--1200}\BibitemShut {NoStop}%
\bibitem [{\citenamefont {Gorbar}\ \emph {et~al.}(2001)\citenamefont {Gorbar},
  \citenamefont {Gusynin},\ and\ \citenamefont {Miransky}}]{GorbarGusynin}%
  \BibitemOpen
  \bibfield  {author} {\bibinfo {author} {\bibfnamefont {E.~V.}\ \bibnamefont
  {Gorbar}}, \bibinfo {author} {\bibfnamefont {V.~P.}\ \bibnamefont
  {Gusynin}},\ and\ \bibinfo {author} {\bibfnamefont {V.~A.}\ \bibnamefont
  {Miransky}},\ }\bibfield  {title} {\bibinfo {title} {{Dynamical chiral
  symmetry breaking on a brane in reduced QED}},\ }\href
  {https://doi.org/10.1103/PhysRevD.64.105028} {\bibfield  {journal} {\bibinfo
  {journal} {Phys. Rev. D}\ }\textbf {\bibinfo {volume} {64}},\ \bibinfo
  {pages} {105028} (\bibinfo {year} {2001})}\BibitemShut {NoStop}%
\bibitem [{\citenamefont {Teber}(2012)}]{Teber_2012}%
  \BibitemOpen
  \bibfield  {author} {\bibinfo {author} {\bibfnamefont {S.}~\bibnamefont
  {Teber}},\ }\bibfield  {title} {\bibinfo {title} {Electromagnetic current
  correlations in reduced quantum electrodynamics},\ }\href
  {http://dx.doi.org/10.1103/PhysRevD.86.025005} {\bibfield  {journal}
  {\bibinfo  {journal} {Physical Review D}\ }\textbf {\bibinfo {volume} {86}},\
  \bibinfo {pages} {025005} (\bibinfo {year} {2012})}\BibitemShut {NoStop}%
\bibitem [{\citenamefont {Hsiao}\ and\ \citenamefont
  {Son}(2017)}]{Son_selfdual}%
  \BibitemOpen
  \bibfield  {author} {\bibinfo {author} {\bibfnamefont {W.-H.}\ \bibnamefont
  {Hsiao}}\ and\ \bibinfo {author} {\bibfnamefont {D.~T.}\ \bibnamefont
  {Son}},\ }\bibfield  {title} {\bibinfo {title} {{Duality and universal
  transport in mixed-dimension electrodynamics}},\ }\href
  {https://doi.org/10.1103/PhysRevB.96.075127} {\bibfield  {journal} {\bibinfo
  {journal} {Phys. Rev. B}\ }\textbf {\bibinfo {volume} {96}},\ \bibinfo
  {pages} {075127} (\bibinfo {year} {2017})}\BibitemShut {NoStop}%
\bibitem [{\citenamefont {Ludwig}\ \emph {et~al.}(1994)\citenamefont {Ludwig},
  \citenamefont {Fisher}, \citenamefont {Shankar},\ and\ \citenamefont
  {Grinstein}}]{Fisher_Hall}%
  \BibitemOpen
  \bibfield  {author} {\bibinfo {author} {\bibfnamefont {A.~W.~W.}\
  \bibnamefont {Ludwig}}, \bibinfo {author} {\bibfnamefont {M.~P.~A.}\
  \bibnamefont {Fisher}}, \bibinfo {author} {\bibfnamefont {R.}~\bibnamefont
  {Shankar}},\ and\ \bibinfo {author} {\bibfnamefont {G.}~\bibnamefont
  {Grinstein}},\ }\bibfield  {title} {\bibinfo {title} {Integer quantum hall
  transition: An alternative approach and exact results},\ }\href
  {https://doi.org/10.1103/PhysRevB.50.7526} {\bibfield  {journal} {\bibinfo
  {journal} {Phys. Rev. B}\ }\textbf {\bibinfo {volume} {50}},\ \bibinfo
  {pages} {7526} (\bibinfo {year} {1994})}\BibitemShut {NoStop}%
\bibitem [{\citenamefont {Herzog}\ \emph {et~al.}(2007)\citenamefont {Herzog},
  \citenamefont {Kovtun}, \citenamefont {Sachdev},\ and\ \citenamefont
  {Son}}]{Mtheory}%
  \BibitemOpen
  \bibfield  {author} {\bibinfo {author} {\bibfnamefont {C.~P.}\ \bibnamefont
  {Herzog}}, \bibinfo {author} {\bibfnamefont {P.}~\bibnamefont {Kovtun}},
  \bibinfo {author} {\bibfnamefont {S.}~\bibnamefont {Sachdev}},\ and\ \bibinfo
  {author} {\bibfnamefont {D.~T.}\ \bibnamefont {Son}},\ }\bibfield  {title}
  {\bibinfo {title} {Quantum critical transport, duality, and m theory},\
  }\href {http://dx.doi.org/10.1103/PhysRevD.75.085020} {\bibfield  {journal}
  {\bibinfo  {journal} {Physical Review D}\ }\textbf {\bibinfo {volume} {75}},\
  \bibinfo {pages} {085020} (\bibinfo {year} {2007})}\BibitemShut {NoStop}%
\bibitem [{\citenamefont {Huh}\ \emph {et~al.}(2013)\citenamefont {Huh},
  \citenamefont {Strack},\ and\ \citenamefont {Sachdev}}]{HuhSachdev}%
  \BibitemOpen
  \bibfield  {author} {\bibinfo {author} {\bibfnamefont {Y.}~\bibnamefont
  {Huh}}, \bibinfo {author} {\bibfnamefont {P.}~\bibnamefont {Strack}},\ and\
  \bibinfo {author} {\bibfnamefont {S.}~\bibnamefont {Sachdev}},\ }\bibfield
  {title} {\bibinfo {title} {{Conserved current correlators of conformal field
  theories in 2+1 dimensions}},\ }\href
  {https://doi.org/10.1103/PhysRevB.88.155109} {\bibfield  {journal} {\bibinfo
  {journal} {Phys. Rev. B}\ }\textbf {\bibinfo {volume} {88}},\ \bibinfo
  {pages} {155109} (\bibinfo {year} {2013})},\ \bibinfo {note} {[Erratum:
  Phys.Rev.B 90, 199902 (2014)]}\BibitemShut {NoStop}%
\bibitem [{\citenamefont {Romatschke}\ and\ \citenamefont
  {S\"appi}(2019)}]{Romatschke:2019qbx}%
  \BibitemOpen
  \bibfield  {author} {\bibinfo {author} {\bibfnamefont {P.}~\bibnamefont
  {Romatschke}}\ and\ \bibinfo {author} {\bibfnamefont {S.}~\bibnamefont
  {S\"appi}},\ }\bibfield  {title} {\bibinfo {title} {{Thermal free energy of
  large Nf QED in 2+1 dimensions from weak to strong coupling}},\ }\href
  {https://doi.org/10.1103/PhysRevD.100.073009} {\bibfield  {journal} {\bibinfo
   {journal} {Phys. Rev. D}\ }\textbf {\bibinfo {volume} {100}},\ \bibinfo
  {pages} {073009} (\bibinfo {year} {2019})}\BibitemShut {NoStop}%
\bibitem [{\citenamefont {Giombi}\ \emph {et~al.}()\citenamefont {Giombi},
  \citenamefont {Tarnopolsky},\ and\ \citenamefont
  {Klebanov}}]{GiombiKlebanov}%
  \BibitemOpen
  \bibfield  {author} {\bibinfo {author} {\bibfnamefont {S.}~\bibnamefont
  {Giombi}}, \bibinfo {author} {\bibfnamefont {G.}~\bibnamefont
  {Tarnopolsky}},\ and\ \bibinfo {author} {\bibfnamefont {I.~R.}\ \bibnamefont
  {Klebanov}},\ }\bibfield  {title} {\bibinfo {title} {{On $C_{J}$ and $C_{T}$
  in Conformal QED}},\ }\href {https://doi.org/10.1007/JHEP08(2016)156}
  {\bibfield  {journal} {\bibinfo  {journal} {JHEP}\ }\textbf {\bibinfo
  {volume} {08}}\bibinfo  {number} { (2016)},\ \bibinfo {pages}
  {156}}\BibitemShut {NoStop}%
\bibitem [{\citenamefont {Leigh}\ and\ \citenamefont
  {Petkou}()}]{Leigh:2003ez}%
  \BibitemOpen
\bibfield  {number} {  }\bibfield  {author} {\bibinfo {author} {\bibfnamefont
  {R.~G.}\ \bibnamefont {Leigh}}\ and\ \bibinfo {author} {\bibfnamefont
  {A.~C.}\ \bibnamefont {Petkou}},\ }\bibfield  {title} {\bibinfo {title}
  {{SL(2,Z) action on three-dimensional CFTs and holography}},\ }\href
  {https://doi.org/10.1088/1126-6708/2003/12/020} {\bibfield  {journal}
  {\bibinfo  {journal} {JHEP}\ }\textbf {\bibinfo {volume} {12}}\bibinfo
  {number} { (2003)},\ \bibinfo {pages} {020}}\BibitemShut {NoStop}%
\bibitem [{\citenamefont {Coleman}(1988)}]{Coleman_Aspects}%
  \BibitemOpen
\bibfield  {number} {  }\bibfield  {author} {\bibinfo {author} {\bibfnamefont
  {S.}~\bibnamefont {Coleman}},\ }\href@noop {} {\emph {\bibinfo {title}
  {Aspects of Symmetry: Selected Erice Lectures}}}\ (\bibinfo  {publisher}
  {Cambridge University Press},\ \bibinfo {address} {Cambridge},\ \bibinfo
  {year} {1988})\BibitemShut {NoStop}%
\bibitem [{\citenamefont {Polyakov}(1987)}]{Polyakov:GFaS}%
  \BibitemOpen
  \bibfield  {author} {\bibinfo {author} {\bibfnamefont {A.~M.}\ \bibnamefont
  {Polyakov}},\ }\href@noop {} {\emph {\bibinfo {title} {{Gauge Fields and
  Strings}}}}\ (\bibinfo  {publisher} {Routledge},\ \bibinfo {address}
  {London},\ \bibinfo {year} {1987})\BibitemShut {NoStop}%
\bibitem [{\citenamefont {Fuini}\ and\ \citenamefont
  {Yaffe}()}]{Fuini:2015hba}%
  \BibitemOpen
  \bibfield  {author} {\bibinfo {author} {\bibfnamefont {J.~F.}\ \bibnamefont
  {Fuini}}\ and\ \bibinfo {author} {\bibfnamefont {L.~G.}\ \bibnamefont
  {Yaffe}},\ }\bibfield  {title} {\bibinfo {title} {{Far-from-equilibrium
  dynamics of a strongly coupled non-Abelian plasma with non-zero charge
  density or external magnetic field}},\ }\href
  {https://doi.org/10.1007/JHEP07(2015)116} {\bibfield  {journal} {\bibinfo
  {journal} {JHEP}\ }\textbf {\bibinfo {volume} {07}}\bibinfo  {number} {
  (2015)},\ \bibinfo {pages} {116}}\BibitemShut {NoStop}%
\bibitem [{\citenamefont {Grozdanov}\ and\ \citenamefont
  {Poovuttikul}()}]{Grozdanov:2017kyl}%
  \BibitemOpen
\bibfield  {number} {  }\bibfield  {author} {\bibinfo {author} {\bibfnamefont
  {S.}~\bibnamefont {Grozdanov}}\ and\ \bibinfo {author} {\bibfnamefont
  {N.}~\bibnamefont {Poovuttikul}},\ }\bibfield  {title} {\bibinfo {title}
  {{Generalised global symmetries in holography: magnetohydrodynamic waves in a
  strongly interacting plasma}},\ }\href
  {https://doi.org/10.1007/JHEP04(2019)141} {\bibfield  {journal} {\bibinfo
  {journal} {JHEP}\ }\textbf {\bibinfo {volume} {04}}\bibinfo  {number} {
  (2019)},\ \bibinfo {pages} {141}}\BibitemShut {NoStop}%
\bibitem [{\citenamefont {Hofman}\ and\ \citenamefont
  {Iqbal}(2018)}]{Hofman:2017vwr}%
  \BibitemOpen
\bibfield  {number} {  }\bibfield  {author} {\bibinfo {author} {\bibfnamefont
  {D.~M.}\ \bibnamefont {Hofman}}\ and\ \bibinfo {author} {\bibfnamefont
  {N.}~\bibnamefont {Iqbal}},\ }\bibfield  {title} {\bibinfo {title}
  {{Generalized global symmetries and holography}},\ }\href
  {https://doi.org/10.21468/SciPostPhys.4.1.005} {\bibfield  {journal}
  {\bibinfo  {journal} {SciPost Phys.}\ }\textbf {\bibinfo {volume} {4}},\
  \bibinfo {pages} {005} (\bibinfo {year} {2018})}\BibitemShut {NoStop}%
\bibitem [{\citenamefont {Grozdanov}\ \emph {et~al.}(2019)\citenamefont
  {Grozdanov}, \citenamefont {Lucas},\ and\ \citenamefont
  {Poovuttikul}}]{Grozdanov:2018fic}%
  \BibitemOpen
  \bibfield  {author} {\bibinfo {author} {\bibfnamefont {S.}~\bibnamefont
  {Grozdanov}}, \bibinfo {author} {\bibfnamefont {A.}~\bibnamefont {Lucas}},\
  and\ \bibinfo {author} {\bibfnamefont {N.}~\bibnamefont {Poovuttikul}},\
  }\bibfield  {title} {\bibinfo {title} {{Holography and hydrodynamics with
  weakly broken symmetries}},\ }\href
  {https://doi.org/10.1103/PhysRevD.99.086012} {\bibfield  {journal} {\bibinfo
  {journal} {Phys. Rev. D}\ }\textbf {\bibinfo {volume} {99}},\ \bibinfo
  {pages} {086012} (\bibinfo {year} {2019})}\BibitemShut {NoStop}%
\bibitem [{\citenamefont {DeWolfe}\ and\ \citenamefont
  {Higginbotham}(2021)}]{DeWolfe:2020uzb}%
  \BibitemOpen
  \bibfield  {author} {\bibinfo {author} {\bibfnamefont {O.}~\bibnamefont
  {DeWolfe}}\ and\ \bibinfo {author} {\bibfnamefont {K.}~\bibnamefont
  {Higginbotham}},\ }\bibfield  {title} {\bibinfo {title} {{Generalized
  symmetries and 2-groups via electromagnetic duality in $AdS/CFT$}},\ }\href
  {https://doi.org/10.1103/PhysRevD.103.026011} {\bibfield  {journal} {\bibinfo
   {journal} {Phys. Rev. D}\ }\textbf {\bibinfo {volume} {103}},\ \bibinfo
  {pages} {026011} (\bibinfo {year} {2021})}\BibitemShut {NoStop}%
\bibitem [{\citenamefont {Skenderis}(2002)}]{Skenderis_holo_renormalization}%
  \BibitemOpen
  \bibfield  {author} {\bibinfo {author} {\bibfnamefont {K.}~\bibnamefont
  {Skenderis}},\ }\bibfield  {title} {\bibinfo {title} {Lecture notes on
  holographic renormalization},\ }\href
  {https://doi.org/10.1088/0264-9381/19/22/306} {\bibfield  {journal} {\bibinfo
   {journal} {Classical and Quantum Gravity}\ }\textbf {\bibinfo {volume}
  {19}},\ \bibinfo {pages} {5849} (\bibinfo {year} {2002})}\BibitemShut
  {NoStop}%
\bibitem [{\citenamefont {Taylor}()}]{Taylor_counterterms}%
  \BibitemOpen
  \bibfield  {author} {\bibinfo {author} {\bibfnamefont {M.}~\bibnamefont
  {Taylor}},\ }\href@noop {} {\bibinfo {title} {{More on counterterms in the
  gravitational action and anomalies}}},\ \Eprint
  {https://arxiv.org/abs/hep-th/0002125} {arXiv:hep-th/0002125} \BibitemShut
  {NoStop}%
\bibitem [{\citenamefont {Frangi}()}]{Giorgio}%
  \BibitemOpen
  \bibfield  {author} {\bibinfo {author} {\bibfnamefont {G.}~\bibnamefont
  {Frangi}},\ }\href@noop {} {\bibinfo {title} {{Geometrisation of Ohm's
  reciprocity relation in a holographic plasma}}},\ \Eprint
  {https://arxiv.org/abs/2406.16124} {arXiv:2406.16124 [hep-th]} \BibitemShut
  {NoStop}%
\bibitem [{\citenamefont {Polonyi}(2014)}]{Polonyi:2014rpa}%
  \BibitemOpen
  \bibfield  {author} {\bibinfo {author} {\bibfnamefont {J.}~\bibnamefont
  {Polonyi}},\ }\bibfield  {title} {\bibinfo {title} {{Classical and quantum
  effective theories}},\ }\href {https://doi.org/10.1103/PhysRevD.90.065010}
  {\bibfield  {journal} {\bibinfo  {journal} {Phys. Rev. D}\ }\textbf {\bibinfo
  {volume} {90}},\ \bibinfo {pages} {065010} (\bibinfo {year}
  {2014})}\BibitemShut {NoStop}%
\bibitem [{\citenamefont {Polonyi}(2015)}]{Polonyi:2015cna}%
  \BibitemOpen
  \bibfield  {author} {\bibinfo {author} {\bibfnamefont {J.}~\bibnamefont
  {Polonyi}},\ }\bibfield  {title} {\bibinfo {title} {{Dissipation and
  decoherence by a homogeneous ideal gas}},\ }\href
  {https://doi.org/10.1103/PhysRevA.92.042111} {\bibfield  {journal} {\bibinfo
  {journal} {Phys. Rev. A}\ }\textbf {\bibinfo {volume} {92}},\ \bibinfo
  {pages} {042111} (\bibinfo {year} {2015})}\BibitemShut {NoStop}%
\bibitem [{\citenamefont {Liu}\ and\ \citenamefont
  {Glorioso}(2018)}]{Glorioso:2018wxw}%
  \BibitemOpen
  \bibfield  {author} {\bibinfo {author} {\bibfnamefont {H.}~\bibnamefont
  {Liu}}\ and\ \bibinfo {author} {\bibfnamefont {P.}~\bibnamefont {Glorioso}},\
  }\bibfield  {title} {\bibinfo {title} {{Lectures on non-equilibrium effective
  field theories and fluctuating hydrodynamics}},\ }\href
  {https://doi.org/10.22323/1.305.0008} {\bibfield  {journal} {\bibinfo
  {journal} {PoS}\ }\textbf {\bibinfo {volume} {TASI2017}},\ \bibinfo {pages}
  {008} (\bibinfo {year} {2018})}\BibitemShut {NoStop}%
\bibitem [{\citenamefont {Grozdanov}\ and\ \citenamefont
  {Polonyi}(2015)}]{Grozdanov:2015nea}%
  \BibitemOpen
  \bibfield  {author} {\bibinfo {author} {\bibfnamefont {S.}~\bibnamefont
  {Grozdanov}}\ and\ \bibinfo {author} {\bibfnamefont {J.}~\bibnamefont
  {Polonyi}},\ }\bibfield  {title} {\bibinfo {title} {{Dynamics of the electric
  current in an ideal electron gas: A sound mode inside the quasiparticles}},\
  }\href {https://doi.org/10.1103/PhysRevD.92.065009} {\bibfield  {journal}
  {\bibinfo  {journal} {Phys. Rev. D}\ }\textbf {\bibinfo {volume} {92}},\
  \bibinfo {pages} {065009} (\bibinfo {year} {2015})}\BibitemShut {NoStop}%
\end{thebibliography}%

\end{document}